\documentclass[reprint,aps,prl]{revtex4-1}
\usepackage{graphicx}
\usepackage{bm}
\usepackage{color}
\usepackage{subfigure}
\usepackage{ulem}
\usepackage{amsmath}
\usepackage{braket}
\usepackage{hyperref}

\begin{document}
\title{The Bose-Einstein Condensate of G-wave Molecules and Its Intrinsic Angular Momentum} 
\author{Tin-Lun Ho}
\affiliation{Department of Physics, The Ohio State University, Columbus, OH 43210, USA} 
\date{\today}
\begin{abstract}
The recent report on the realization of a Bose-Einstein condensate of G-wave molecule made up of bound pairs of Cesium bosons is a surprise. These molecules are created at the G-wave resonance at 19.87G, where the severe three-body loss usually associated with these resonance are found to be  reduced significantly when the density of the gas is reduced in a quasi 2D setting.
 The G-wave molecules produced through this resonance have  non-zero angular momentum projections, resulting in the first BEC with a macroscopic intrinsic angular momentum. Here, we show that this intrinsic angular momentum will lead to many new quantum effects. They include a splitting of collective modes in the absence of vortices,  an orientation dependent energy shift due to the moment of inertia of the molecules, and a contribution to angular momentum in non-uniform magnetic fields different from that of the Berry phase current. The intrinsic angular momentum also provides a way to probe the half vortices, excitations that are unique to molecular condensates of bosons. The   wavefunction giving rise to the intrinsic angular momentum can also be mapped out from the noise correlation. 
\end{abstract}
\maketitle

Superfluids are media where quantum coherence is displayed on a  macroscopic scale. Bose-Einstein condensates (BEC) are special kinds of superfluids. 
Even at densities  one hundred thousand times thinner than air, they still exhibit the same quantum coherence as liquid Helium. In addition, the quantum states of the atom in a BEC can be manipulated easily because of their rich atomic energy level structure. 
The combination of quantum state control and macroscopic quantum coherence makes quantum gases attractive systems for developing quantum technology and sensors. As demonstrated dramatically in superfluid $^{3}$He, the number of macroscopic quantum phenomena increases significantly even when the  condensate wavefunction gains a few more degrees of freedom. Compared to atoms, molecules have  many more degrees of freedom and far more complex energy level structures. They can therefore afford a  greater range of quantum state control, and BECs with  richer quantum phenomena. 

Soon after the success of creating fermion bound pairs using a wide Feshbach resonance\cite{Jin, Grimm, Jin2, Ketterle}, there have been continual efforts to create molecular BEC's of fermions and bosons pairs with non-zero angular momentum. However, such efforts have faced a serious challenge, as non s-wave resonances often has severe three-body loss associated, which depletes the quantum gas rapidly. This makes the recent report on the realization of a BEC of G-wave molecules (G-wave BEC for short) particularly surprising and exciting\cite{Chin}. By reducing the density of the gas in a quasi 2D trap, Cheng Chin's group at Chicago has found that the three-body loss is substantially reduced at low temperatures for a G-wave resonances of $^{133}$Cs. From the equation of state deduced from the density profile, they concluded that the molecular gas is a BEC. While more direct confirmations of BEC  (such as momentum distribution measurements) are needed, the increased stability of the molecular gas at low temperatures shows a major hurtle against BEC formation has been reduced. In this paper, we shall discuss the unique properties of BECs of boson pairs with non-zero angular momentum,  ($\langle {\bf L}\rangle \neq 0$). These phenomena,  absent in atomic BEC,  will help identify the presence of the G-wave condensate. 

The G-wave BEC also provides an opportunity to address some intriguing issues in condensed matter. It has been predicted that the transition between the molecular and atomic BECs must be first order\cite{Stoof,Leo}. This can be seen from their $2\pi$ vortices. In an atomic BEC, each boson will acquire a phase change of $2\pi$ around the vortex core, whereas in the molecular BEC, it is the molecule that acquires a $2\pi$ phase change. This means each boson carries only half of a vortex, and hence cannot be connected smoothly to that in an atomic BEC. With the realization of the G-wave BEC, this prediction can be tested.  The G-wave BEC can also shed lights on the longstanding problem of the intrinsic angular momentum of superfluid $^{3}$He-A. There have been many studies in the last few decades that led to vastly different answers\cite{Anderson, Ish, Cross, Stone, Machida}. The issue is about the amount of edge current that mixes with the angular momentum of the Cooper pair. This issue has not been unsettled experimentally as it is difficult to separate different sources of angular momentum. For the G-wave BEC, the molecules are of atomic size, much less than the interparticle distance.  As a result, the contribution of the edge current is negligible, and the intrinsic angular momentum of the pairs can be measured precisely. On the other hand, one can induce edge currents by tightening up the confining trap, or by bringing the system close to resonance to increase the pair size. The study of the angular momentum of the G-wave BEC can therefore shed lights on the decades old $^{3}$He-A problem. 

{\em G-wave molecule and its intrinsic angular momentum:} 
The G-wave molecule studied in ref.\cite{Chin} is a tightly bound pair of Cs bosons in the hyperfine spin state $\chi^{}_{4,4}=|F=4, F_z=4\rangle$, and orbital angular momentum state $|L=4, L_{z}=2\rangle$, both with quantization axis along the external magnetic field ${\bf B}=B\hat{\bf z}$. Its wavefunction is 
\begin{equation}
 \Psi({\bf R},  {\bf r}) = C({\bf R}) F({\bf r}) \chi^{}_{4,4}, \,\,\,\, F({\bf r}) =  f(r)  Y_{4,2}(\hat{\bf r}), 
\label{F} \end{equation}
where ${\bf R}$ and ${\bf r}$ are the center of mass and relative coordinates of the pair,  $Y_{4,2}(\hat{\bf r})$ is the spherical harmonic,  $f$ is the normalized radial wavefunction, and $C$ is the wavefunction of the center of mass.   This molecule is obtained by sweeping through the narrow Feshbach resonance at 19.87 G, which has a width of 11 mG. The experiments are performed away from the resonance in the range of $18.2G < B < 19.5G$, where $f$ is found to have a width $d_{0}=5.3$nm. The system is confined in a 2D trap with inter-molecule spacing $d \sim 1.6 \times 10^{-5}$cm. 
The immediate consequence of the molecular state (Eq.(\ref{F})) is that it has an intrinsic angle momentum 
\begin{equation}
    \vec{\cal L} = \langle {\bf r}\times {\bf p}\rangle = 2\hbar \hat{\bf B}.
\end{equation}
Hence, a vortex free G-wave BEC with $N$ molecules will have  angular momentum $2\hbar N$, the same as a doubly quantized vortex. 

For S-wave molecules, $Y= 1$, Eq.(\ref{F}) reduces to the function $C({\bf R})$ on the length scale  larger than $d_{M}$.  Consequently,  $C({\bf R})$ becomes the condensate wavefunction in a many-body system, similar to that of atomic BEC. The situation is different for molecules with non-trivial angular function, as  shrinking its range to zero does not remove its angular function. As we shall see,  this angular function when coupled to  external perturbations will lead to effective potentials for the center of mass function $C({\bf R})$, which we now illustrate. 

The hamiltonian of a molecule in free space is $h(1,2) = \sum_{i=1,2}{\bf p}^{2}_{i}/2M - \overline{\mu} {\bf B}\cdot {\bf F} + {\cal V} = {\bf P}^2/M + h_{o}$, where $F={\bf F}_{1}+{\bf F}_{2}$ is the total hyperfine spin, $\overline{\mu}$ and $M$ are the magnetic moment and the mass of boson (Cs), and ${\cal V}$ is the interaction between bosons. ${\cal V}$ has a dipolar form as a result of a strong second order spin-orbit coupling\cite{Paul}.  ${\bf P}$ is the total momentum. $h_{o}$ is the hamiltonian in the center of mass frame and is a function of relative coordinate  ${\bf r}= {\bf r}_{1} - {\bf r}_{2}$ only. The molecular state Eq.(\ref{F}) is an eigenstate of $h_{o}$ with energy $\epsilon_{M}$, $h_{o}G = \epsilon_{M} G$. 
 
In a rotating trap with angular velocity $\bm{\Omega}$ aligned with the magnetic field ${\bf B}$, the hamiltonian is $K =h(1,2) +  \sum_{i=1,2} [   u({\bf r}_{i})  - \bm{\Omega}\cdot {\bf r}_{i}\times {\bf p}_{i} ]$, where $u({\bf r})$ is the trapping potential. Since all terms in $K$ except ${\cal V}$ vary slowly on the size of the molecule, we can expand them around the center of mass ${\bf R}= ({\bf r}_{1} + {\bf r}_{2})/2$, and average over the molecular wavefunction $F$ in Eq.(\ref{F}). The resulting energy is 
\begin{equation}
\langle K \rangle_{\Psi} = \int_{\bf R}  C^{\ast}( {\bf R}) {\cal K}({\bf R}, {\bf P}) C({\bf R})
\end{equation}
\begin{eqnarray}
     {\cal K}({\bf R}, {\bf P}) = \langle K\rangle_{F}
    = \frac{{\bf P}^2}{4M} -\bm{\Omega}\cdot{\bf R}\times {\bf P} + 2U({\bf R}) + \epsilon_{M}  \hspace{0.2in}  \nonumber \\  
   -  \bm{\Omega}\cdot \vec{\cal L}  + \Delta E, \,\,\,\,\,\,\,\,
   \Delta E=  \frac{1}{2} I_{ab} (\omega^2)_{ab},   \hspace{0.5in}
\label{K} \end{eqnarray}
where $I_{ab} = 2M\langle r_{a}r_{b}/4 \rangle_{F} $ is the moment of inertia tensor of the molecule, 
and $(\omega^2)_{ab} \equiv \nabla_{a}\nabla_{b}U({\bf R})/M$. Defining 
${\rm Tr} I \equiv (2M) (d^{2}_{M}/4)$, where $d_{M}$ is the size of the molecule, the angular function $Y_{4,2}$ implies the following  dyadic form for $I_{ab}$, 
\begin{equation}
\hat{I} = (M d^{2}_{M}/2) (23 \hat{\bf x}\hat{\bf x} + 23 \hat{\bf y}\hat{\bf y} + 31\hat{\bf z}\hat{\bf z})/77.   \label{I}  
\end{equation}
We shall write $d_{M}= \lambda d_{0}$.  In the range of magnetic field in Ref.\cite{Chin}, we have $d_{M}=d_{0}$, i.e  $\lambda=1$. For simple potential scattering, $\lambda$ will increase considerably close to resonance. For the Cs G-wave molecule, there are also s-wave contributions to the wavefunction within the width of the resonance. How much increase $\lambda$ will increase can be found in future experiments.  

Both  $\vec{\cal L}$ and  $I_{ab}$ are intrinsic properties of the G-wave molecule. Their effects are less prominent in the normal phase due to the fluctuations of the orientation of the molecular spin,  but are magnified in a BEC. Because the small size of the molecule, the effects of $I$ and $\Delta E$ are very weak. However, their presence can be found as follows.

\noindent (A) Transitions to higher rotational states: If one turns on a small oscillating potential, $\delta u = {\rm cos}\omega t V({\bf r})$, it follows from the derivation of Eq.(\ref{K}) that   
a potential  ${\rm cos}\omega t ({\bf r}\cdot \nabla /2)^2 V({\bf R}) $ will be generated for the molecule. This is a second order tensor in ${\bf r}$ that will cause transitions to higher rotational states when $\omega $ matches the energy difference, which is typical of order of MHz\cite{Paul}.

\noindent (B) Energy shift in deep optical trap: Eq.(\ref{K}) and (\ref{I}) show that in a harmonic trap $V_{har} = \frac{1}{2} M\sum_{i}\omega_{i}^2 x_{i}^2$, $i=x,y,z$, the energy shifts is 
$\Delta E = \sum_{i} (\hbar\omega_{i}/2) \alpha_{i}(d_{M}/L_{i})^2$, where $L_{i}= \sqrt{\hbar/M\omega_{i}}$, and $\alpha_x. \alpha_{y}, \alpha_z = (23, 23, 31)/77$. For an optical lattice 
constructed from a 1064nm laser with lattice constant $a=532nm$, a deep lattice of 20 to 30 recoils will create a harmonic well on each lattice site with oscillator length 10 to 20nm, which is a few times more than the size of molecule (5.3nm). The shift $\Delta E$ will then be a few percent of the frequencies $\omega_{i}$, which is measurable. 


{\em Many-body wavefunction:} The wavefunction of $N$ molecules is  a function $2N$  bosons invariant under particle exchange. Since the system favors {\em tightly} bound pairs,  the normalized ground state will then be a symmetrized product of pairs
\begin{eqnarray}
 {\cal G}( 1,2, .. 2N) =  \bigg( \Psi(1,2)\Psi(3,4) .. \Psi(2N-1, 2N) +  \bigg. \nonumber \\
\bigg.   + {\rm other \, \, distinct \,\, pair \, \, products } \bigg) \frac{1}{\sqrt{(2N-1)!!}}.   \hspace{0.3in}
\label{GG} \end{eqnarray}
where (1,2 ...) stands for the positions $({\bf x}_{1}, {\bf x}_{2}, ...)$ of the bosons, and  $\Psi(1,2)$ is the normalized molecular wavefunction Eq.(\ref{F}),  
$\int_{1,2} |\Psi(1,2)|^2=1$. Since the size of the bound pair is much smaller than inter-molecule distance $d$, the overlap of different pair products is exponentially small for typical configurations and hence can be ignored.  Within the same approximation, the ground state Eq.(\ref{GG}) has the second quantized form, 
\begin{equation}     |{\cal G}\rangle = \frac{1}{\sqrt{N!}}{\cal O}^{\dagger}|0\rangle, \,\,\,\,\, {\cal O}^{\dagger} = \frac{1}{\sqrt{2}}\int_{\bf r, r'}
\Psi({\bf r}, {\bf r'})\psi^{\dagger}({\bf r})\psi^{\dagger}({\bf r'}). 
\label{calG} \end{equation}
For  single particle operators ${\cal O}_{i}$ and  short range  two body operators ${\cal W}_{ij}$, we have 
 \begin{eqnarray}
     \sum_{i=1}^{2N} \langle  {\cal O}_{i} \rangle_{\cal G} 
     = N({\cal O}_{1} + {\cal O}_{2}\rangle_{\Psi(1, 2)}+ (...) , 
 \label{single}  \\
 \sum_{1\leq i <j\leq 2N} \langle  {\cal W}_{ij} \rangle_{\cal G} = N \langle {\cal W}_{1, 2}  
    \rangle_{\Psi(1, 2)} + (...) .  
\label{W}  \end{eqnarray}
 where $(...)$ means those much smaller exchange contributions. 
With Eq.(\ref{F}), (\ref{K}), (\ref{single}), (\ref{W}), and the assumption that the molecules interact with a contact interaction $g$, 
the energy functional of the molecular BEC is 
\begin{equation}
    E[\Phi]= \int_{\bf R} 
    \Phi^{\ast}({\bf R}) {\cal K}({\bf R}, {\bf P}) \Phi({\bf R}) + \frac{g}{2}\int |\Phi({\bf R})|^4,  \hspace{0.2in}
\label{GP}\end{equation}
where $\Phi({\bf R})= \sqrt{N} C({\bf R})$ is the wavefunction of the molecular condensate,  $\int_{\bf R} |\Phi({\bf R})|^2=N$, and  $g$ can be extracted from the solution of the four-body problem.   

{\em The intrinsic angular momentum}: Eq.(\ref{single}) implies that the total angular momentum of a molecular BEC with $N$ molecules is 
\begin{equation}
    \langle {\bf L} \rangle_{\cal G} = \int_{\bf R} \Phi^{\ast}({\bf R}) ({\bf R}\times {\bf P})  \Phi({\bf R})+ (2\hbar)N \hat{\bf B}.  
\end{equation}
Even the system has no vortices (hence vanishing first term), it  has an angular momentum $2N\hbar$. In atomic BEC, it was shown in ref.\cite{Stringari} that the angular momentum of a condensate will lead to splittings of its collective modes. This result was established with two steps. The first is a set of  rigorous relations relating the transition moments $ m_{p}^{\pm}= \int^{\infty}_{0}{\rm d}
E \left[ S_{+}(E)\pm  S_{-}(E)\right]E^{p}$ to the angular momentum and the moment of inertia of the system, where $S_{\pm}(E)=\sum_{n}|\langle n|F_{\pm}|0\rangle|^2 \delta(E-\hbar \omega)_{n0}$, and  $F_{\pm}= \sum_{i}(x\pm i y)^2_{i}$ (or $F_{\pm}= \sum_{i}(x\pm i y)_{i} z_{i}$) for the quadrupole mode (or the dipole mode).
 The second is the assumption that these moments will be exhausted by the the modes $\omega_{\pm}$ in the hydrodynamic limit, which means $S_{\pm}(E) = \sigma^{\pm}\delta(E-\hbar \omega_{\pm})$ with strengths $\sigma^{\pm}$. 
 This validity of this assumption was discussed in \cite{Stringari2}, and the prediction in ref.\cite{Stringari}  was verified in the experiments on atomic BEC\cite{Dalibard}. Repeating these derivations for the molecular BEC using Eq.(\ref{single}), one finds for the quadrupole mode 
\begin{equation}
    (\omega_{+}- \omega_{-})_{quad} = \frac{2}{M} \frac{\langle L^{total}_{z}\rangle_{\cal G}}{\sum_{\alpha=1}^{N}
    \langle {\bf R}^{2}_{\perp \alpha} + {\bf r}^{2}_{\perp \alpha}/4\rangle_{\cal G} }, 
\label{quad} \end{equation}
\begin{equation}
    (\omega_{+}- \omega_{-})_{dipole} = \frac{2}{M} \frac{\langle L^{total}_{z}\rangle_{\cal G}}{\sum_{\alpha=1}^{N}
    \langle {\bf R}^{2}_{\perp \alpha} + 2({\bf R}\cdot \hat{\bf z})^2+ \Delta_{\alpha}^{2}\rangle_{\cal G} }, 
\label{dipole} \end{equation}
where $\alpha$ labels the molecules, and $\Delta^{2}_{\alpha} 
= {\bf r}^{2}_{\perp \alpha}/4 + z^{2}_{\alpha}/2$. Away from the resonance, the terms ${\bf r}_{\perp \alpha}^{2}$ and $\Delta^{2}_{\alpha}$ in Eq.(\ref{quad}) and (\ref{dipole})  can be ignored because of the small size of the molecule. 

\begin{figure}
\centering
\includegraphics[width=3.2in, height=2.5in]{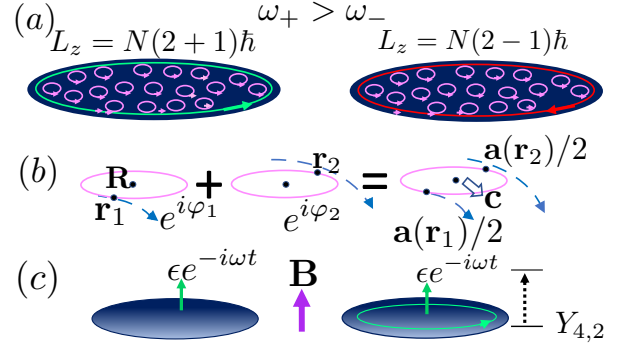}
\caption{(a) 
The total angular momentum of  a $+2\pi$ and a $-2\pi$-vortex (left and right figure): The intrinsic angular momentum $2\hbar N$ is along $\hat{\bf z}$. $N$ is the number of molecules.  The total angular momentum of the $ +2\pi$ and $-2\pi$ vortex are $L_{z} = 3N\hbar$ and $N\hbar$ respectively. Since $L_{z}$ has the same sign in both cases,  so will be the  frequency shifts $(\omega_{+}-\omega_{-})$ of their quadrupolar modes.  If the intrinsic angular momentum was absent, then $L_{z}$ (and hence the frequency shifts) of 
$+$ and $-2\pi$ vortices will have different signs. 
(b) The half vortex is a equal superposition of $e^{i\varphi_{1}}$ and $e^{i\varphi_{2}}$. The vortex field of the boson at ${\bf r}_{i}$ is ${\bf a}({\bf r}_{i})/2$. Their difference is ${\bf c}= ({\bf r}\cdot \nabla) {\bf a}({\bf R})/4$, represented by an unfilled arrow.
(c) If the trap has a small oscillatory rotation $e^{-i\omega t}$, it will generate in a vortex state (right figure) 
a term $- \epsilon e^{-i\omega t}  \hat{\bf z}\times {\bf r}\cdot {\bf c}$ in the two-body hamiltonian, which will cause transition to a higher molecular rotational state when $\omega$ matches the energy difference.  No such transitions will occur in the absence of vortices, (left figure).  } 
\label{fig1}
\end{figure}

{\em Vortices:}
In atomic BEC, a vortex corresponds augmenting every boson with a phase factor $\zeta= e^{i\varphi}$, where  $\varphi$ is the azimuthal angle of its position ${\bf R}$. The gradient of the phase angle, referred to as the  ``vortex field", is 
\begin{equation}
    {\bf a}({\bf R}) = \nabla \varphi({\bf R}) = \frac{\hat{\bf z}\times {\bf R}}
    {|\hat{\bf z}\times {\bf R}|^2}. 
\label{a} \end{equation}
However, applying this procedure to the two-boson molecule can only create  a double quantized vortex, as the  factor  $e^{i(\varphi({\bf r}_{1}) + \varphi({\bf r}_{2}))}$ generates a $4\pi$ phase change around the vortex core . 
A phase function that satisfies Bose statistics and describes a singly quantized vortex is  
\begin{equation}
    \zeta({\bf r}_{1}, {\bf r}_{2}) = \frac{e^{i\varphi_{1}} + e^{i\varphi_{2}}}{|e^{i\varphi_{1}} + e^{i\varphi_{2}}|} 
    =  \frac{e^{i(\varphi_{1} + \varphi_{2})/2} {\rm cos}(\varphi_{1}-\varphi_{2})}{|{\rm cos}(\varphi_{1}-\varphi_{2})|}, \hspace{0.2in}
\label{zeta}\end{equation}
where $\varphi_{i}= \varphi({\bf r}_{i})$.  Because of the symmetric superposition,  Eq.(\ref{zeta}) can be viewed as a  $2\pi$-vortex shared by two bosons, each of which effectively carries half of a $2\pi$-vortex. Moreover, 
${\bf a}_{i}({\bf r}_{i}) = \nabla_{i} {\rm Arg} (\zeta) = {\bf a}({\bf r}_{i})/2$, $i=1,2$. The total angular momentum of a $2\pi q$ vortex $(\zeta^{{\pm} q})$ is 
\begin{equation}
    L_{z} = \hbar (2 \pm q)  N, 
\label{LL} \end{equation}
where the first term is due to the intrinsic angular momentum, and $q$ is an integer. See Figure 1(a).
In passing, we point out that  a $2\pi$-vortex  of a BEC of $n$-boson bound is
\begin{equation}
  \eta(1,2, ..n) = \bigg( \prod_{n\geq i>j\geq 1}
\zeta_{ij} \bigg)^{2/n(n-1)},
\end{equation}
with ${\bf a}_{i}({\bf r}_{i}) = \nabla_{i}{\rm Arg} (\eta)= {\bf a}({\bf r}_{i})/n$. 

It has been pointed out that the half vortices of boson pair condensates implies a a first order phase transition between atomic and molecular BEC\cite{Stoof, Leo}. 
The phase function in Eq.(\ref{zeta}) suggests another way to probe the presence of the half vortices.  
We had shown that in a rotating trap, the intrinsic angular momentum gives rise to a term 
 $h_{I} = -\Omega \hat{\bf z}\cdot {\bf r}\times {\bf p}$ in relative coordinates. See Eq.(\ref{K}).  
In the presence of a vortex $\zeta^{q}$, the momentum of each boson shifts  as ${\bf p}_{i}\rightarrow {\bf p}_{i} + {\bf a}_{i}({\bf r}_{i})$. This lead to a shift in the relative momentum ${\bf p}\rightarrow {\bf p}+{\bf c}$, where ${\bf c}({\bf R}, {\bf r})= ({\bf a}_{1}({\bf r}_{1})
  -{\bf a}_{2}({\bf r}_{2}))/2
  = q({\bf r}\cdot \nabla){\bf a}({\bf R})/4$, and hence  a shift 
  \begin{equation}
    \delta h_{I} = - q\Omega (\hat{\bf z}\times {\bf r}) \cdot \left(  {\bf r}\cdot \nabla \right){\bf a}({\bf R})/4. 
  \end{equation}
A small oscillatory rotation $\Omega = \epsilon {\rm cos}\omega t$  can excite the molecule to other higher rotational states, as in our previous discussion for the molecular moment of inertia.  See Figure 1(b) and 1(c).   Combining this transition (independent of the sign of $q$) with the measurement on the splitting of collective modes (Eq.(\ref{LL})) for vortices with $q=\pm 1$ will make a strong case for the phase function Eq.(\ref{zeta}).  
\begin{figure}
\centering
\includegraphics[width=3.2in, height=2.5in]{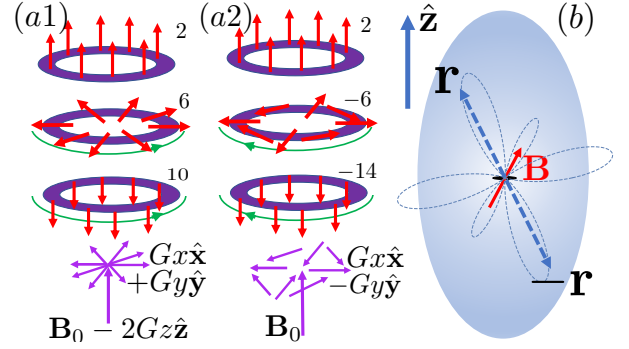}
\caption{(a) and (b): We consider a sum of an uniform magnetic field ${\bf B}_{0}$ and a quadrupolar field $B_{1}= G(x\hat{\bf x}+ y\hat{\bf y} - 2z\hat{\bf z})$ (Figure (a)), or 
$B_{1}= G(x\hat{\bf x}- y\hat{\bf y})$, (Figure (b)).  The BEC is trapped in the annulus region in xy-plane (at $z=0$). We then have $\hat{\bf B}(R,\varphi, z=0) = {\rm sin}\theta ({\rm cos}\varphi \hat{\bf x} + {\rm sin}\varphi \hat{\bf y}) + {\rm cos}\theta\hat{\bf z}$, with ${\rm tan}\theta = GR/B_{0}$. The variation of the angular momentum quantization axis gives rise to a Berry phase current (green color), and an angular momentum $ 6N(1-{\rm cos}\theta(R))$, (the first term of Eq.(\ref{Lz}), where $N$ is the number of molecules.   The intrinsic angular momenta of the molecules follow their local fields. Their total contribution along $\hat{\bf z}$ is $2N\hat{\bf z}\cdot \hat{\bf B}$. The numerals next to the figure (a) and (b) are the angular momenta per molecule. (c) Schematics of the free expansion of the G-wave molecular BEC after the interaction and the trap is suddenly turned off. Initial, the BEC (depicted as a short black line) is located at the center. The direction of the magnetic field in the initial state is indicated by a purple arrow. Because of the tight confinement of the initial state along $z$, the cloud expands predominantly in that direction. Its shape is a convolution of the condensate wavefunction with the molecular wavefunction, Eq.(\ref{ex}). The noise correlation at ${\bf r}$ and $-{\bf r}$ is given by $|Y_{4,2}(\hat{\bf r}; \hat{\bf B}))|^2$, represented by the dashed curve. Measuring the noise correlation in the high density regions of the expanding cloud for different directions $\hat{\bf B}$ will map out angular function $|Y_{4,2}|^2$.
} 
\label{fig2}
\end{figure}

{\em Angular momentum in non-unform magnetic field}: 
The Zeeman energy of a molecule in a non-uniform magnetic field is $-\sum_{i=1,2}{\bf B}({\bf r}_{i})\cdot {\bf F}_{i}$. Expanding it about the center of mass, we have $-{\bf B}({\bf R})\cdot ({\bf F}_{1} + {\bf F}_{2}) + ({\bf r}\cdot \nabla/2){\bf B}\cdot ({\bf F}_{1} - {\bf F}_{2})$. The second term can be ignored as it vanishes when averaged over the pair wavefunction. The first term is a uniform rotation of the two spins. Since the interaction ${\cal V}$ that forms the molecule has a dipole form, it is invariant under simultaneous rotation of spin and  relative coordinates\cite{Paul}. The molecular state (Eq.(\ref{F})) now  becomes  $\Psi({\bf R}, {\bf r}) = C({\bf R}) f(r) |\chi ({\bf R})\rangle$, where  $|\chi ({\bf R})\rangle$ is the angular momentum eigenstate along the magnetic field, 
$|\chi ({\bf R})\rangle = |F=4,{\bf F}\cdot \hat{\bf B}({\bf R})=4\rangle \otimes |L=4,{\bf L}\cdot \hat{\bf B}({\bf R})=2\rangle$. We can take $C({\bf R})$ to be real, as its phase can be absorbed in  $|\chi ({\bf R})\rangle$.  Applying the procedure in Eq.(\ref{K}) and (\ref{single}), we find that
\begin{equation}
   \langle {\bf L}\rangle_{\cal G} = \int {\rm d}{\bf R} \bigg(  {\bf R} \times 
    {\bf g}({\bf R}) + 2\hbar \hat{\bf B}({\bf R}) 
    \bigg) |\Phi({\bf R})|^2
\end{equation}
where ${\bf g}({\bf r}) = -i\hbar \langle \chi ({\bf R})| \nabla | \chi ({\bf R}) \rangle $ is the momentum current generated by the non-uniform spin state, (i.e. the Berry phase current). 
For magnetic fields of the form 
${\bf B}(R, \varphi, z) = {\rm cos}\theta \hat{\bf z} + 
{\rm sin}\theta({\rm cos}\varphi \hat{\bf x} \pm 
{\rm sin}\varphi \hat{\bf y})$, where  $(R, \varphi, z)$ are the cylindrical coordinates and $\theta$ is only a function of $R$, we have 
\begin{eqnarray}
    {\bf g}(R, \varphi, z) = \pm  \, 6\hbar \frac{1- {\rm cos}\theta}{R}\hat{\varphi}, 
\label{g}  \hspace{1.4in} \\
    L_{z} = 2\pi \hbar \int_{0}^{\infty}  \bigg(  2 
   {\rm cos} \theta  \pm 6 (1-{\rm cos}\theta) \bigg) |\Phi(R)|^2 R{\rm d} R. \hspace{0.4in}
\label{Lz} \end{eqnarray}
The ``6" in Eq.(\ref{g}) and (\ref{Lz}) is the total spin projection $F_z + L_z$.  
The magnetic field mentioned above can be generated by superposing a uniform field ${\bf B}_{0}$ with a quadrupolar field 
${\bf B}_{1} = G (x\hat{\bf x} + y\hat{\bf y} - 2 z\hat{\bf z})$ or ${\bf B}_{1} = G (x\hat{\bf x} - y\hat{\bf y})$ , corresponding to the $(+)$ and $(-)$ sign in Eq.(\ref{Lz})), where $G$ is the magnetic field gradient. If the BEC is confined in a 2D annulus as in Figure 2 where the field ${\bf B}$ is essentially uniform along the radial direction, then we have 
$L_z/N = (0, 6, 10)\hbar$ and 
$L_z/N=( 0, -6, - 14)\hbar$ when  $\theta = (0, \pi/2, \pi)$ for the $+$ and $-$ cases in Eq.(\ref{Lz}). See Figure 2(a) and 2(b).  

{\em Mapping out the angular function}: Finally, we show that the coherence of the BEC will enable one to map out the molecular wave function. Consider suddenly changing the magnetic field to the value of zero boson scattering length,  and turning off the trap simultaneously. In the center of mass frame of a molecule, the bosons will fly off in $({\bf k}, -{\bf k})$-pair in the distribution $|Y_{4,2}(\hat{\bf k}; \hat{\bf B})|^2$. At the same time,  the center of mass wavefunction will expand predominantly along the tightly confined direction. To determine the molecular angular function, we need to remove the center of mass motion.

Prior to expansion, the gas (of size $L$) was trapped in an anisotropic trap centered around the origin. 
It was shown\cite{Demler} that after the gas has expanded to a size much greater then $L$,  the density $n({\bf r}, t)= \langle \psi^{\dagger}({\bf r},t)\psi^{}({\bf r},t) \rangle_{\cal G}$ gives  the momentum distribution of the initial state $|{\cal G}\rangle$, i.e. $n({\bf r},t) = \beta(t) \langle a^{\dagger}_{\bf k} a^{}_{\bf k} \rangle_{\cal G}$, with ${\bf k} = M{\bf r}/\hbar t$, and $\beta$ is a function of $t$ only.  With the ground state Eq.(\ref{GG}), we have 
\begin{equation} 
\langle a^{\dagger}_{\bf k} a^{}_{\bf k} \rangle_{\cal G} = 2N \sum_{\bf p} \big| \widetilde{\Psi}({\bf k}, {\bf p}) \big|^2 =  
\sum_{\bf p} |\widetilde{\Phi}({\bf k}-{\bf p})|^2 |\widetilde{F}({\bf p})|^2.  
\label{ex} \end{equation}
where $\widetilde{\Psi}({\bf k}, {\bf p})$, $\widetilde{\Phi}({\bf k})$, and $\widetilde{F}({\bf k})$  
are the Fourier transform of $\Psi({\bf r}_{1}, {\bf r}_{2})$, $\Phi({\bf R})= \sqrt{N} C({\bf R})$ and $F({\bf r})$ in Eq.(\ref{F}) respectively. For a 2D trap, the condensate wavefunction $\Phi({\bf R})$ has a pancake shape in xy-plane, hence $\widetilde{\Phi}({\bf k})$ is elongated along $z$. 
For atomic BEC, the  momentum distribution is simply the Fourier transform of condensate wavefunction, 
$\langle a^{\dagger}_{\bf k} a^{}_{\bf k} \rangle_{\cal G}= |\widetilde{\Phi}({\bf k})|^2$. 
 For molecular condensates, Eq.(\ref{ex}) shows $\langle a^{\dagger}_{\bf k} a^{}_{\bf k} \rangle_{\cal G}$ is a convolution of the condensate wavefunction $\Phi$ with the molecular wavefunction $F$. Separating out these two functions from the expanded density is not a simple task. 
 
On the other hand, the center of mass motion can be filtered out if we consider the noise correlation at at opposite points $({\bf r}, -{\bf r})$, 
$G({\bf r},t;  -{\bf r},t) \equiv \langle \Delta n({\bf r}, t) \Delta n({\bf r}', t)\rangle  - \langle \Delta n({\bf r}, t)\rangle \langle \Delta n({\bf r}', t)\rangle $. It is straightforward to show that
\begin{eqnarray}
   \langle \Delta n({\bf r}, t) \Delta n(-{\bf r}, t)\rangle = N |\widetilde{F}({\bf k}, -{\bf k})|^2  \nonumber \\ 
   = N \widetilde{f}(k)^2 
   |Y_{4,2}(\hat{\bf k}; \hat{\bf B})|^2, \,\,\,\,\,\,  \hat{\bf k}= \hat{\bf r}. \hspace{0.2in}
\end{eqnarray}
where $\widetilde{f}(k) = 4 \pi \int^{\infty}_{0} j_{4}(kr) f(r) r {\rm d}r$, and $j_{\ell}(kr)$ is the spherical Bessel function.  
Even though the gas expands predominantly along $z$, the entire 
the function $|Y_{4,2}({\bf k}; \hat{\bf B})|^2$ can still be mapped out by repeating the measurements for different ${\bf B}$. See Figure 2(c). 

{\em Concluding Remarks}: We have discussed the signatures of the intrinsic angular momentum and the half vortices of the G-wave molecular BEC. The range of properties we have discussed also apply to other molecular BECs with non-zero angular momentum. So far we have only discussed the phenomena far from resonance where the size of molecules is much smaller than the distance between them.  The physics near unitarity will be of great interests. Apart from new type of unitarity physics, these studies will also help us explore the first order transition between atomic and molecular BEC, and further our understanding of the intrinsic angular momentum of $^{3}$He-A. In current experiments\cite{Chin}, the  angular momentum of Cs G-wave molecules are frozen along the external magnetic field. It is  like $^{3}$He-A on a substrate where the angular momentum of the Cooper pairs are frozen along the surface normal. In one could create a stable molecular gas with intrinsic angular momentum that can orient freely, the BEC of this gas would be a fascinating superfluid indeed.

Acknowledgments: I thank Cheng Chin and Martin  Zwierlein for stimulating discussions, and Cheng for the estimates of optical lattice parameters needed to probe the energy shift of the G-wave molecule in confinement. I also thank Paul Julienne for discussions on the Cs G-wave molecules. The work is supported by the MURI Grant FP054294-D.


\begin{thebibliography}{99}
\bibitem{Chin} Zhendong Zhang, Liangchao Chen, Kaixuan Yao, Cheng Chin, Atomic Bose-Einstein condensate to molecular Bose-Einstein condensate transition, arXiv:2006.15297. 
\bibitem{Jin} C. A. Regal, M. Greiner, and D. S. Jin, Emergence of a molecular Bose–Einstein condensate from a Fermi gas, NATURE Vol 42, 537 (2003). 
\bibitem{Grimm} S. Jochim, M. Bartenstein, A. Altmeyer, G. Hendl, S. Riedl, C. Chin, J. Hecker Denschlag, R. Grimm, Bose-Einstein Condensation of Molecules, Science Vol 302, 2101, (2003). 
\bibitem{Jin2} C. A. Regal, M. Greiner, and D. S. Jin, Observation of Resonance Condensation of Fermionic Atom Pairs, Physical Review Letters {\bf 92}, 040403 (2004).
\bibitem{Ketterle} M. W. Zwierlein, C. A. Stan, C. H. Schunck, S. M. F. Raupach, A. J. Kerman, and W. Ketterle, Condensation of Pairs of Fermionic Atoms near a Feshbach Resonance, 
Phys. Rev. Lett. 92, 120403 (2004). 
\bibitem{Stoof} M.W.J. Romans, R.A. Duine, S. Sachdev, and H. T. C.  Stoof, Quantum phase transition in atomic Bose gas with a Feshbach resonance. Physical Review Letters {\bf 93}, 020405 (2004)
\bibitem{Leo} L. Radzihovsky, J. Park, and P.B.  Weichman, Superfluid transition in bosonic atom-molecule mixture near a Feshbach resonance. Physical Review Letters {\bf 92}, 160402 (2004). 
\bibitem{Anderson} P. W. Anderson and P. Morel, Generalized Bardeen-Cooper-Schrieffer States of the Proposed Low-Temperature Phase of Liquid He$^{3}$, Physical Review {\bf 123}, 1911, (1961). 
\bibitem{Ish} M. Ishikawa, Orbital Angular Momentum of Anisotropic Superfluid, Progress of Theoretical Physics {\bf 57} 1836, (1977). 
\bibitem{Cross} M.C. Cross, A Generalized Ginzburg-Landau Approach to the Superfluidity of Helium 3, Journal of Low Temperature Physics {\bf 21}, 525, (1975). 
\bibitem{Stone} Michael Stone and Rahul Roy, Edge modes, edge currents and gauge invariance in $p_{x}+ip_{y}$ superfluids and superconductors, Physical Review B {\bf 69} 184511, (2004).
\bibitem{Machida} Y. Tsutsumi, T, Mizushima, M. Ichioka, K. Machida, On Intrinsic Angular Momentum due to Edge Mass Current for Superfluid $^{3}$He A-Phase, Journal of Physics: Conference Series {\bf 400}  012076 (2012)
\bibitem{Paul} Martin Berninger, Alessandro Zenesini, Bo Huang, Walter Harm, Hanns-Christoph Naigerl, Francesca Ferlaino, Rudolf Grimm, Paul S. Julienne, and Jeremy M. Hutson, Feshbach resonances, weakly bound melecular states, and coupled-channel potentials for cesium at high magnetic fields, Physical Review A87, 032517 (2013).
\bibitem{Stringari} Francesca Zambeli and Sandro Stringari, Quantized Vortices and Collective Oscillation of a Trapped Bose-Einstein Condensate, Physical Review Letters 81, 1754 (1998).
\bibitem{Stringari2} S. Stringari, Sum rules for density and particle excitations in Bose superfluids,  Physical Review B46, 2974 (1992).
\bibitem{Dalibard} F. Chevy, K.W. Madison, and J. Dalibard, Measurement of the Angular Momentum of a Rotating Bose-Einstein Condensate, Physical Review Letters 85, 2223 (2000) 
\bibitem{Demler} Ehud Altman, Eugene Demler, and Mikhail D. Lukin, 
Probing many-body states of ultracold atoms via noise correlations, Physical Review {\bf A 70}, 013603 (2004)
\end{thebibliography}
\end{document}